\documentclass{sigchi-ext}
\usepackage[T1]{fontenc}
\usepackage{textcomp}
\usepackage[scaled=.92]{helvet} 
\usepackage{graphicx} 
\usepackage{balance}  
\usepackage{booktabs} 
\usepackage{ccicons}  
\usepackage{ragged2e} 
\usepackage{todonotes}

\def\plaintitle{Situated Case Studies for a Human-CenteredDesign of Explanation User Interfaces} \def\plainauthor{Claudia M\"uller-Birn, Katrin Glinka, Peter S\"orries, Michael Tebbe, Susanne Michl}

\def\plainkeywords{Human-Centered XAI; case study design; critical reflection}

\title{Situated Case Studies for a Human-Centered Design of Explanation User Interfaces}

\numberofauthors{5}
\author{%
    \alignauthor{%
    \textbf{Claudia Müller-Birn}\\
    \affaddr{Freie Universität Berlin \\ Human-Centered Computing} \\
    \affaddr{Berlin, Germany} \\
    \email{clmb@inf.fu-berlin.de} }\alignauthor{%
    
     \textbf{Michael Tebbe}\\
    \affaddr{Freie Universität Berlin \\ Human-Centered Computing} \\
    \affaddr{Berlin, Germany} \\
    \email{michael.tebbe@fu-berlin.de} } \vfil \alignauthor{%
    
    \textbf{Katrin Glinka}\\
    \affaddr{Freie Universität Berlin \\ Human-Centered Computing} \\
    \affaddr{Berlin, Germany} \\
    \email{katrin.glinka@fu-berlin.de} }\alignauthor{%
    
    \textbf{Susanne Michl}\\
    \affaddr{Charité Medical Humanities and \\ Ethics in Medicine}\\
    \affaddr{Berlin, Germany} \\
    \email{susanne.michl@charite.de} } \vfil \alignauthor{%

    \textbf{Peter S\"orries}\\
    \affaddr{Freie Universität Berlin \\ Human-Centered Computing} \\
    \affaddr{Berlin, Germany} \\
    \email{peter.soerries@fu-berlin.de} \\ }\alignauthor{%
} }

\definecolor{linkColor}{RGB}{6,125,233}
\hypersetup{%
  pdftitle={\plaintitle},
  pdfauthor={\plainauthor},
  pdfkeywords={\plainkeywords},
  bookmarksnumbered,
  pdfstartview={FitH},
  colorlinks,
  citecolor=black,
  filecolor=black,
  linkcolor=black,
  urlcolor=linkColor,
  breaklinks=true
}

\begin{document}

\setcopyright{none}

\maketitle

\RaggedRight{} 

\begin{abstract}
Researchers and practitioners increasingly consider a human-centered perspective in the design of machine learning-based applications, especially in the context of Explainable Artificial Intelligence (XAI). However, clear methodological guidance in this context is still missing because each new situation seems to require a new setup, which also creates different methodological challenges. Existing case study collections in XAI inspired us; therefore, we propose a similar collection of case studies for human-centered XAI that can provide methodological guidance or inspiration for others. We want to showcase our idea in this workshop by describing three case studies from our research. These case studies are selected to highlight how apparently small differences require a different set of methods and considerations. With this workshop contribution, we would like to engage in a discussion on how such a collection of case studies can provide a methodological guidance and critical reflection.
\end{abstract}




\section{Introduction}
While machine learning (ML)-based software is increasingly used in all parts of our society, researchers and practitioners emphasize the importance of a human-centered perspective when designing these systems (e.g.~\cite{baumer2017toward, gilliesHumanCentredMachineLearning2016, Kogan2020:humancentereddatascience, benjamin_materializing_2019}). In this context, Explainable AI (XAI) techniques are increasingly used. The former, i.e., specific algorithms such as LIME~\cite{ribeiro_why_2016} or SHAP~\cite{Lundberg:2017tc}, aim to support human understanding and sense-making of the results of ML-based software by providing explanations. The XAI techniques can be categorized into explanation methods (e.g. local, global, example-based or counterfactual)~\cite{liao_questioning_2020}; therefore, the scope and type of the explanations provided differ as do their content and representation. The overall goal of XAI techniques is to provide interpretability and transparency~\cite{carvalho2019MLinterpretability,molnar2020interpretable}, since research has shown that the degree to which the ML results can be interpreted by explanations can enhance user understanding, which, in turn, leads to more trust~\cite{stumpf2007toward}. Therefore, explanations need to be carefully designed in \textit{Explanation User Interfaces} (Explanation UI) for their context of use~\cite{tomsett2018interpretable, liao_questioning_2020}.
%

The characteristics of Explanation UIs needed are currently under research (e.g. \cite{cheng_explaining_2019, liao_questioning_2020, kaur2020interpreting, jesus2021can}). Previous research has shown that study participants have had difficulties to understand the underlying conceptional model of the XAI techniques provided. Nonetheless, the explanations provided conveyed trust in the XAI techniques. At the same time, the 'ease of use' reduced the critically thinking of the study participants~\cite{kaur2020interpreting}. Efforts to improve evaluation approaches of XAI techniques still disconnect the Explanation UI from the context of use~\cite{jesus2021can}, even though insights from social science emphasize the subjectivity of explanations~\cite{miller_explanation_2019}. This highlights the need to adapt them to the context and audience~\cite{miller_explanation_2019}. Explanations can have different purposes that relate to specific (intelligibility) questions, such as situation-specific information in the form of \textit{why} or \textit{why not} questions or \textit{what if} questions that allow people to probe the model~\cite{liao_questioning_2020, lim2009assessing}. 

In our research, we missed a methodological guidance ourselves\footnote{There are valuable exceptions to this claim, such as the research by Ehsan et al.~\cite{ehsan2021expanding}, Eiband et al.~\cite{eiband2018bringing} and Baumer~\cite{baumer2017toward}.}, which motivates our proposal for assembling \textit{Situated Case Studies}. Our proposal is inspired by existing initiatives such as the XAI stories\footnote{More information:  \url{https://pbiecek.github.io/xai_stories/}.} that focus on technical aspects of XAI and the ''Princeton Dialogues on AI and Ethics Case Studies''\footnote{More information: \url{https://aiethics.princeton.edu/case-studies/}.}, and extends approaches such as the Explainability Fact Sheet~\cite{sokol2020explainability}. Different from these approaches, we focus on a design perspective. In order to emphasize the notion of situatedness, we suggest a reflective approach to design. A situated approach does not attempt ''to establish one correct understanding and set of metrics'' when designing interactions but instead studies the ''local, situated practices of users''~\cite{harrison2011making}. This, in turn, requires us to reconsider how much of an intended outcome lies in the control of a designer. As Sch\"on~\cite{schoen1983reflective} highlights, the actual effect of a design might differ from the designer's initial intention. The designer, therefore, has to enter into a ''conversation'' with a situation, by a chosen methodological approach. This should create a dynamic in which ''the situation 'talks back', and [the designer] responds to the situation's back-talk''~\cite[p.79]{schoen1983reflective}. Sch\"on describes this conversation with a situation as reflective (ibid.). Sengers et al.~\cite{sengers2005reflective} emphasize this as a need for the \textit{critical} reflection of both users and designers in the design process. 


In the following, we introduce three case studies from our ongoing research. We consider each of these case studies as ''situated'' and as such in need of a particular methodological approach in order to enter into such a ''conversation.'' Human-centered design (HCD) comprises a number of dimensions~\cite{KlingStar1998:HCD} that can be used to define angles for reflection that should be considered in the design of Explanation UIs: (1) Who does the usage of the ML-based system affect? (\textit{stakeholder}); (2) whose purposes are served in the design process and whose not? (\textit{purpose}); and (3) how will the design of the ML-based system impact people’s experience? What unintended consequences might result from the design and the deployment of the ML-based system? (\textit{context}). By taking this approach, we expect to gain a deeper understanding of the specific characteristics of each case study, which will later allow us to reflect better on the choice of methodology. Our goal is to subsequently reintegrate our methodological choices to inform future design and research in the area of Explanation UIs.

\section{Situated Case Studies}

In the following, we briefly describe three selected case studies in which we apply a HCD approach in the context of XAI. These research studies aim to enable a closer, more effective and transparent collaboration between humans and machines, especially with non-technical experts or lay users.

\subsection{Case 1: Explaining Privacy-Preserving Machine Learning}

The first case study is situated in the medical domain and focuses on the value-oriented data donation of patients. This research aims to balance the trade-off between the need for unrestricted data in individualized medicine, on the one hand, and the protection of patients' personal data, on the other. There is a particular urgency (especially due to the GDPR\footnote{The General Data Protection Regulation (GDPR) is a regulation in EU law on data protection and privacy in the European Union (EU).}) of applying privacy-preserving ML, i.e. here differential privacy (DP)~\cite{papernot2018scalable}, in the clinical context. We explore in this case study how existing possibilities and limitations of DP can be explained to the patients (stakeholders) in such a way that it enables them to make informed decisions (purpose) about their data donation (context).
At the beginning of the project, we envisioned a typical HCD process, assuming that the problem was well-defined: We would need to investigate the \textit{explainability needs}\footnote{According to Liao et al. \cite{liao_questioning_2020}, \textit{explainability needs} ''represent categories of prototypical questions users may ask to understand AI.''} of our stakeholders while the purpose and context of use were clearly outlined. Reality taught us otherwise. We experienced that even in our interdisciplinary project team (consisting of experts in medicine, machine learning, security and HCI) many questions existed regarding the ML pipeline (e.g. how does DP affect the accuracy of ML predictions?) and the explanation of DP (e.g. how can different privacy levels be explained and realized in DP?). Besides patients, clinical researchers need to understand the capabilities and limitations of this technology; therefore, Explanation UIs are needed for lay users, i.e., patients, and non-technical experts, i.e., clinical researchers. Thus, we had to take a step back and extended our design approach by considering the clinical researchers as well. We extend Wolf's suggestion of using \textit{explainability scenarios}~\cite{wolf_explainability_2019} as a resource for designing for interpretability and also use them as a resource for reflecting on the effects of our designs. Instead of focusing on the explanation methods, i.e. what can be explained, a scenario concentrates on the \textit{explainability needs}, i.e. which \textit{explainability needs} align with what explanation method? We have so far defined three scenarios which can be used to engage with our different stakeholders. 

Each scenario provides an Explanation UI in a different context of use. We plan to use these scenarios as resource for reflection in co-creation workshops. While our focus in this use case still lies on the patients as designated users of our Explanation UIs for data donation, our process so far has shown that we also need to factor in the clinical researchers who need a deeper understanding of privacy preserving ML to raise their acceptance in applying this technology.



\subsection{Case 2: Explaining Interactive Clustering Results}

The second case study originates from the area of digital media studies, where ML techniques are increasingly used to handle large-scale data in qualitative research settings (e.g.~\cite{chen2018using, smith2018closing}). This case study is motivated by the need to scale up qualitative interpretive research~\cite{buzzanell2018interpretive} of textual comments from YouTube with a ML-based data analysis pipeline (purpose). Based on a close interdisciplinary collaboration between media studies and HCI researchers (stakeholder), we built a text analysis pipeline in which -- after pre-processing the data -- the semantic similarity between sentences is computed based on embeddings generated by a pre-trained language model. Uniform Manifold Approximation and Projection~\cite{mcinnes2018umap} is then used for dimensionality reduction and k-medoids for clustering similar comments. The resulting basic clustering visualization is used in an iterative process of (re-)labeling the data. During the latter, the major sense-making with the data takes place (context). 
It turned out that this approach of using the pipeline to refine the model iteratively was challenging. The situation was primarily caused by the cluster visualization, which represented the ML pipeline results. The results often did not align with the mental model of the non-technical expert, which is a typical problem in human-AI interaction (cf.~\cite{amershi2019guidelines}). At the same time, the HCI researcher realized that the visualization was taken for given when presenting preliminary results, and the media researcher did not critically reflect on the limitations of the quantitative approach (e.g. existing bias in the word embeddings). Baumer echoes this observation from his research and calls for allowing a non-technical audience to critically interrogate the data and explore alternative perspectives~\cite{baumer2017toward}. 

In the design process, we experienced these unintended consequences of our ML pipeline and realized the need for an \textit{Explanation UI} that supports a critical reflection. However, we had difficulties to collect and concretely describe implicit \textit{explainability needs} that the media researcher had not yet become aware of. More specifically, we realized that we require a better understanding of how explanations might influence (or redirect) non-technical experts in their research process. Thus, we decided to design an \textit{Explanation UI} which augments the existing clustering visualization. This interface allows users to, for example, assess the clustering results by exploring the comments contained or relevant features for the word embeddings. Inspired by research from Hohman~\cite{Hohman:Gamut}, we plan to use this \textit{Explanation UI} as a 'technology probe' and will conduct contextual inquiries to identify the implicit \textit{explainability needs}. We will then translate the implicit needs into explanations and explore, finally, how these available explanations impact the sense-making process of non-technical experts.

\subsection{Case 3: Explanations in Narrative-based Decision-Making}

 The third case study relates to an ongoing research collaboration with a medical ethicist. As has already been mentioned, ML is increasingly used in clinical settings to support decision-making (e.g.~\cite{papernot2018scalable}). However, in complex situations, there is a need for a holistic perspective of the patient, their (family) situation, preferences and moral concepts of what a good life represents for them. What is best for the patient does not only depend on external evidence, but has to be found out in each individual situation in a narrative structured decision-making process. Physicians might use ML-based systems for understanding possible therapy outcomes in advance. In the meeting, however, they would then present 'their' interpretation of the ML result. However, in clinical practice, the medical ethicist envisions a more active involvement of 'the AI'~\footnote{We adopted the term ''AI'' for two reasons. First, the medical ethicist wants the technical system to get its ''own voice'' in the meeting, and second, the ''materiality'' of the technical systems is not yet defined and part of the design process.}.
In this case study, we aim to sketch out an adaptive ML-based system that allows meeting participants to actively engage, dispute, or collaborate with ''the AI'' through explanations as part of the narrative-based decision-making process. Thus, the ML-based system continuously adapts in its capability by providing possible therapy scenarios based on the provided data. Insight into this is crucial in order to understand what factors will impact how (and if) ''the AI'' can contribute to the process and how this might impact the decision-making process. We assume that \textit{Explanation UIs} that incorporate conversational models are especially suitable in this context~\cite{miller_explanation_2019}.

One challenging aspect of this case study is to envision concrete situations and interactions of how physicians and clinical staff can engage with 'the AI' (cf.~\cite{yang2020-Re-Examining-Human-AI-Interaction}). We have started the contextualization by using the medical ethicist's manifold ethnographic experiences as a basis for detailing existing situations within the narrative-based decision-making process where 'the AI' participates. However, this case study is highly exploratory and the impact of using such an 'AI' in this context is rather ambiguous. The level of uncertainty within the design process is especially high compared to the other case studies discussed. Taking this into account and inspired by the research of Dunne and Raby~\cite{dunne2013speculative}, we will use speculative design to imagine alternative futures of how such an 'AI' materializes in the context given. These futures can be probable, plausible, possible and preferable~\cite{dunne2013speculative} and provide a lens to our stakeholders to envision better how such an 'AI' could participate in their present decision-making. We plan to use \textit{Speculative Enactments} as an approach to do speculative design research with participants~\cite{Elsden2017_SpeculativeEnactments}.



\section{Conclusion}

All three case studies include various stakeholder groups, purposes and contexts. While applying our HCD perspective, we were forced to critically reflect on each setup. In our discussions, we used the case studies to derive scenarios that take different perspectives on the design problem space and, thus, reveal consequences of different types~\cite{Carroll:2000gb}. Understanding the particularity of each use case allows us to adapt the HCD methods used accordingly. 
Each situated case study introduces a different challenge. In the first case study, we undermine the role of clinical researchers as stakeholders in the design process. In the second case study, we realized that the purpose of the design needed to be extended, and in the last case study, the context of application is uncertain. In each case, we follow a certain methodological path. Since all case studies are ongoing, we cannot evaluate our methodological choices yet. In any case, we suggest that in the context of human-centered XAI, these case studies can inform and inspire other researchers to reflect on their particular application context, the purpose of each XAI approach, the intended users and the Explanation UI designed. By expanding additional case studies, we hope to contribute to ongoing efforts for the systematic engagement and reflective use of HCD methods to explain ML-based systems.

\section{Acknowledgements}
We thank the reviewers for their insightful comments. This work is supported by the German Research Foundation (EXC 2025: Matters of Activity. Image Space Material) and the Federal Ministry of Education and Research (grant 16SV8463: WerteRadar).

\balance{} 

\bibliographystyle{SIGCHI-Reference-Format}
\bibliography{references}

\end{document}